\documentclass{article}
\usepackage [a4paper,left=2cm,right=2cm,head=1cm,foot=1cm,top=1cm,bottom=2cm] {geometry}
\usepackage{graphicx}
\usepackage{times}
\usepackage[round]{natbib}
\begin{document}

\title{Auto-reverse nuclear migration in bipolar mammalian cells on 
micropatterned surfaces}

\author{B. Szab\'o $^1$, Zs. K\"ornyei $^2$, J. Z\'ach $^3$, D. Selmeczi $^3$, G. Cs\'ucs $^4$, A. Czir\'ok $^3$ and T. Vicsek $^{1,3}$}

\maketitle

$^1$ Research Group for Biological Physics, HAS, Budapest, P\'azm\'any P\'eter s\'et\'any 1/A, 1117 Hungary.

$^2$ Institute of Experimental Medicine, Budapest, Szigony u. 43, Hungary.

$^3$ Department of Biological Physics, E\"otv\"os University, Budapest, P\'azm\'any P\'eter s\'et\'any 1/A, 1117 Hungary. 

Phone: +36 1 372 2795, Fax: +36 1 372 2757

$^4$ Laboratory for Biomechanics, ETH Z\"urich.

\paragraph{Running title:} Auto-reverse nuclear migration in mammalian cells.
\paragraph{Abbreviations:} AMP-PNP, 5'-adenylylimidodiphosphate; MLCK, myosin light-chain kinase; 
MT, microtubule; MTOC microtubule organizing center.
\newpage
\section*{Abstract}
A novel assay based on micropatterning and time-lapse microscopy has been
developed for the study of nuclear migration dynamics in cultured mammalian
cells.  When cultured on 10-20 $\mu$m wide adhesive stripes, the motility of C6
glioma and primary mouse fibroblast cells is diminished. Nevertheless, nuclei
perform an unexpected auto-reverse motion: when a migrating nucleus approaches
the leading edge, it decelerates, changes the direction of motion and
accelerates to move toward the other end of the elongated cell.  During this
process cells show signs of polarization closely following the direction of
nuclear movement.  The observed nuclear movement requires a functioning
microtubular system, as revealed by experiments disrupting the main
cytoskeletal components with specific drugs.  On the basis of our results
we argue that auto-reverse nuclear migration is due to forces determined by the
interplay of microtubule dynamics and the changing position of the microtubule
organizing center as the nucleus reaches the leading edge.  Our assay
recapitulates specific features of nuclear migration (cell polarization,
oscillatory nuclear movement), while allows the systematic study of a large
number of individual cells. In particular, our experiments yielded the first
direct evidence of reversive nuclear motion in mammalian cells, induced by
attachment constraints.

\newpage
\section*{Introduction}

Nuclear migration is an essential aspect of a number of cellular and
developmental processes. Correct positioning of the nucleus within a eucaryotic
cell can have various functions \citep{nuclearpositioning,
Morris00, Morris03}. For example, in budding yeast nuclear migration into the
bud neck is required for correct segregation. In fission yeast a
complex, oscillatory motion has been observed and interpreted with its relation
to meiosis \citep{yeastnucloscill}. Similar microtubule-dependent nuclear 
oscillations were reported in mutant budding yeast cells associated with a 
prolonged mitotic arrest \citep{nucloscillbuddingyeast}
In a number of non-vertebrate species the directed
movements of the male and female pronuclei are also essential in zygote
formation \citep{nuclearpositioning}.
Migration of nuclei to the cortex is the
prerequisite for normal cellularization in the syntitial Drosophila embryo (Foe
and Alberts, 1983).

Much less is known, however, about the role of and the mechanism by which the
nucleus migrates in mammalian cells. The appropriate location of the nucleus
can be relevant in several cell types: For instance, in the tightly packed
pseudostratified epithelia nuclei of the elongated cells are arranged to avoid
each other.  
In the pseudostratified ventricular epithelium of the developing and adult
brain, nuclei of neuronal progenitor cells and radial glial cells oscillate
between the ventricular and cortical surfaces
\citep{VZneuronmigration,AGMM88,NFWDK01}. During this process, termed
`interkinetic nuclear migration', the position of the nucleus systematically
depends upon the cell cycle phase and progenitor cells divide only at the
ventricular surface. Although the mechanism of
this remarkable phenomenon is still not clear \citep{INM,Frade02};
maintaining a strongly polarized, elongated morhoplogy seems 
to be associated with the consistent work of the apparatus responsible for 
moving the nucleus \citep{FWC00}.

Chemically micropatterned surfaces offer a nice and elegant way to manipulate cell shape and to a certain extent also cell function \citep{singhvietal, chenetal2, chenetal, craigheadetal}. We have also shown previously that a method based on microcontact printing is well suited to manipulate cell-shape in a controlled way \citep{csucs}. In order to combine the advantages of time-lapse microscopy, the potential of
{\it in vitro} manipulations and the necessary condition of the cell's
bipolarity we designed an {\it in vitro nuclear migration assay} made of
micropatterned 10 to 20 micron wide, cell adhesive stripes separated from each
other by attachment restricting stripes. The effects of such a confined
geometry are twofold: cells assume an elongated morphology, while their
locomotion is inhibited.  As a result, we have a device in which motion of
the nuclei can be followed in great detail over a long period of time. In turn,
this enables us to characterize the nuclear motion both qualitatively and
quantitatively. 
Our main observation is that cells under such conditions often exhibit a
puzzling oscillatory motion of the nucleus. This process is seen in various
mammalian cell cultures and characterized by continuous deceleration, change in
the direction of motion and acceleration to move towards the other end of the
elongated cell.  

The questions we address in this work are concerned with the
biophysical (forces, dynamics) and the cell biological (cytoskeletal proteins
involved) background  of this reversive nuclear migration. One motivation for
this is the appreciation of the fact that in addition to being a tiny
biochemical factory, our cells contain a number of microscopic machines
performing mechanical work. With great advances in nanotechnology, there is an
increasing potential and need in the quantitative understanding of this
essential -- forces versus motion -- aspect of the cellular machinery. 
\newpage
\section*{Materials and Methods}

\paragraph{Cell Cultures.}

Cultures of C6, U87 gliomas, and 3T3 fibroblast cell lines were grown in Minimum Essential Medium (MEM;
Sigma) supplemented with 10\% fetal calf serum (FCS; Gibco), 4 mM glutamine and
40 $\mu$g/ml gentamycin in humidified air atmosphere containing 5\% CO$_2$, at 37
$^{\circ}$C. Primary fibroblasts were isolated from the bodies of 13 day old
mouse embryos and were grown to confluency in DMEM (Sigma) containing 10\% FCS.
Prior to the time-lapse experiments cells were passaged over the micropatterned
surfaces of 35 mm plastic petri dishes (Greiner)  at a density of
$1-1.5\cdot10^4$ cell/cm$^2$. In some experiments the stripe pattern was
printed on a 25 mm glass coverslip, which was then placed into a petri dish. 

\paragraph{Microscopy.}

Time-lapse recordings were performed on a computer-controlled Leica DM IRB
inverted microscope equipped with 10x, 20x objectives and an Olympus Camedia
4040z digital camera.  Cell cultures were kept at 37 $^{\circ}$C in humidified
5\% CO$_2$ atmosphere within a custom-made incubator attached to the microscope
stage \citep{hebal00,CRRL02}.  Phase contrast images were acquired in every 5
min for 2-3 days.  

\paragraph{Micropatterning.}

Chemically structured substrates containing 10-20 $\mu$m wide stripes
permitting and restricting cell attachment were prepared by microcontact
printing. Adhesive surfaces were coated with either 40 $\mu$g/ml fibronectin
with 20 $\mu$g/ml Alexa488-labeled fibrinogen (Molecular Probes), or 50
$\mu$g/ml FITC-labeled poly-l-lysine (SIGMA).  Blocking of protein/cell
attachment was achieved by a subsequent covering the surfaces with
poly-l-lysine-g-poly(ethylene glycol) (PLL-PEG) \citep{HMV01}.  For further
details see \citep{csucs}.

\paragraph{Drugs.}

Specific drugs (SIGMA) were used to affect various cytoskeletal proteins.
Microtubule (MT) disruption was induced by 20-100 nM vinblastine, MT dynamics was shifted
towards polymerization by 10-100 nM taxol \citep{taxol,MTdrugs}. 
Cytoplasmic dyneins were inhibited
by 5 $\mu$M sodium-orthovanadate \citep{MBW00}, 
kinesins by 100 $\mu$M 5'-adenylylimidodiphosphate (AMP-PNP) \citep{algakinezin}.  
F-actin was influenced by 100-500 nM cytochalasin-D \citep{cytochalasinD}.
Myosin-II activity was inhibited by the myosin light-chain kinase (MLCK) inhibitor ML-7 at 
concentrations of 1 and 10 $\mu$M \citep{ML-7}.  Reagents were added
to the culture medium after one day long time-lapse recording in normal medium.
Drug effectiveness was tested by standard viability assays \citep{MTT}. Briefly, C6 cells 
treated for 24 hour with various concentrations of drugs were incubated with 
3-(4,5-dimethylthiazol-2-yl)-2,5-diphenyltetrazolium bromide (MTT, 25 mg/ml). 2 hours 
later cells and formazan crystals were dissolved in acidic (0.08 M HCl) isopropanol
and transferred onto 96 well plates. Optical density (OD) was determined at a 
measuring wavelenght of 570 nm against 630 nm as reference using an SLT 210 ELISA-reader. 
We administered the drugs in non-toxic concentrations in the time-lapse experiments. The use of vanadate in cell culture is not straightforward because the permeability of the cell membrane to vanadate can be problematic. We used vanadate in a concentration of 5 $\mu$M , as a dose, which still resulted in a reduction of overall cell viability. Time-lapse recordings did not show alterations in the rate of cell decay, cellular morphology or in the nuclear motility of
cells upon treatment with 5 $\mu$M vanadate.

\paragraph{Immunocytochemistry.}

Cells grown on micropatterned glass coverslips were fixed immediately following
the termination of the time-lapse recording (4\% paraformaldehyde in PBS for 20
min at  room temperature ). Membranes of fixed cells were permeabilized by
treatment with Triton X-100 (5 min, 0,1\% v/v in PBS). Non-specific antibody
binding was blocked by incubation with 5\% FCS in PBS at room temperature, for
1 hour. Antibodies to $\alpha$-tubulin (mouse; ExBio, Praha) and
$\gamma$-tubulin (rabbit; Sigma) were used at dilutions of 1 to 1000-5000,
respectively. Secondary antibodies to $\alpha$-tubulin were Cy3 (1:3000,
Jackson) or Alexa488 labeled anti-mouse IgG-s (1:1000, Molecular Probes).
$\gamma$-tubulin was visualised by 1.5 hour incubation with biotin-conjugated
anti-rabbit IgG (1:1000, Vector) followed by 1 hour incubation with fluorescent
avidin-TRITC (1:750, Sigma).  The immunolabeled sample was compared to the last
frame of the obtained recording. By locating the same cells in both images, we
were able to identify the relationship between nuclear motion and the
distribution of labeled protein complexes. Thus, each immunostained frame
contained an assembly of cells with nuclei stopped at various stages of their
motion.

\paragraph{Data processing and analysis.}

Cell nucleus positions together with the two process ends were determined
manually in each time-lapse frame using a custom-written software.  The
procedure resulted in the nucleus position data $x_i(t)$, which denotes the
location of the nucleus of cell $i$ along the stripe at a time point $t$.  The
duration of each nuclear migration period was determined as the elapsed time
between the peaks of $x_i(t)$.  
Velocity of the nucleus was calculated from its net displacement in a 30 
minute-long time interval as $v_i(t)=[x_i(t+\tau/2)-x_i(t-\tau/2)]/\tau$, where
$\tau=30$min.
The nuclear migration activity during a time period $T$ was characterized by
the average magnitude of nuclear velocities as 
$V_T=\langle \vert v_i(t) \vert \rangle_{i,t\in T}$, 
where the average $\langle \rangle_{i,t\in T}$ is calculated over all nuclei 
$i$ which were oscillating within the time period $T$.
Drug effects were characterized by comparing nuclear migration activity of the
first day (control) to that of the second day (drug exposed) as 
$V_{D2}/V_{D1}$.
To establish the statistical significance of the differences, velocity
data was sampled at 1h intervals. The resulting velocity magnitudes
(absolute values), characterizing the cell population before and after the
drug exposure, were compared by two-sample t-tests.

\paragraph{Online supplemental material.}

Video 1 displays the motility of C6 cells on the surface of 20 $\mu$m wide 
Alexa488-labeled fibronectin-fibrinogen stripes. Video 2
shows the reversive nuclear motion in a single C6 cell on a similar stripe. Frame size: 270x60 $\mu$m$^2$. Video 3 (frame size: 400x115 $\mu$m$^2$), 4 (frame size: 250x50 $\mu$m$^2$), and 5 (frame size: 630x480 $\mu$m$^2$) show the effect of 30 nM taxol (administered at 16 h), 20 nM vinblastine (administered at 9 h), and 500 nM cytochalasin D (administered at 12 h) respectively on auto-reverse nuclear migration in C6 cells cultured on 20 $\mu$m wide stripes.  
  
\newpage
\section*{Results}

\paragraph{Nuclear migration assay on micropatterned surfaces.}

Various cell types were cultured on micropatterned substrates, consisting of
alternating stripes permitting (fibronectin-fibrinogen or poly-l-lysine) and restricting
(PLL-PEG) cell attachment.  As Fig.~1 demonstrates, under these conditions
cells assume a highly elongated, bipolar morphology. Cell cultures, grown on
the micropatterned substrates, were recorded for two days with computer
controlled microscopy.  The recordings show that while net cell movements
diminish, locomotory activity remains present: Both processes of the bipolar
cells still exhibit -- intermittently -- a leading edge-like dynamics with
membrane ruffling and filopodium activity (Video 1). 

Surprisingly, cell nuclei often engage in an oscillatory movement.  As shown in
Fig.~2, when a migrating nucleus approaches one of the leading edges, it reverses direction and continues to move in the opposite direction without a lag period (Video 2).
During direction reversals the nuclei do not rotate, but retain orientation
(data not shown).  Nuclei were often observed to perform multiple (up to 15)
such cycles between cell divisions.  The reversive nuclear motion is attachment
constraint specific. It was exhibited in $32\pm7\%$ of C6 cells when
cultured on 20 $\mu$m wide stripes (data from 4 independent cultures). In
non-micropatterned control cultures nuclear oscillation was found solely in
elongated cells, never in flattened cells. Characteristic width of bipolar C6 cells is between 10 and 20 $\mu$m, which implies that on 20 $\mu$m or wider stripes bipolar cells do not respond to the full line width. They still show reversive nuclear motion.

\paragraph{Cell polarization.} 

Cells with a migrating nucleus also show signs of polarization. Usually the
front edge, toward which the nucleus moves, exhibits more membrane ruffling
activity. At the same time the trailing edge narrows and occasionally detaches
(Fig.~2). This polarization reverses together with the directionality of
nuclear motion.  Sister cell nuclei tend to oscillate with opposite
polarization after mitosis,  especially on the very narrow (10 $\mu$m) stripes
(Fig.~3).

\paragraph{Reversive nuclear motion in various cell types.}

The extent of attachment constraint-induced nuclear oscillation is cell
type specific. We found the phenomenon most prominently in C6 cells, 
where the average period length  of the oscillation was $4.6 \pm 0.4$ h 
(Fig. 4a).  In primary fibroblast
cells the nucleus reverses direction further from the leading edges, reducing thereby
their average period length to $1.8 \pm 0.25$ h (Fig. 4b).  3T3 nuclei are also
motile, but usually fail to reverse direction. Finally, U87 cells exhibit a
less elongated morphology and display less pronounced nuclear motility.  A substantial
variability was found among individual cells even within the same culture.  The
population standard deviation of period lengths is approximately $40\%$,
i.e., 1.9 h and 0.6 h for C6 cells and primary fibroblasts, respectively.

\paragraph{Temporal resolution of nuclear motion.}

To better characterize the dynamics of reversive nuclear motion, cell nuclei
and process ends were tracked on consecutive images.  The resulting position
data (Fig. 5) demonstrate that reversals in nuclear motion direction are
coupled to the destabilization and partial detachment of the new trailing
processes. Velocities, calculated from changes in nuclear positions, show that
nuclei slow down, reverse direction and then accelerate in a smooth, continuous manner.
Thus, the temporal changes in positions and velocities give no evidence for that
the gradual deceleration and reversing direction could be attributed to distinct mechanisms.

Despite the differences in period lengths, the average speed (i.e., the average
absolute value of the velocity) of the nuclei was found to be similar in both
C6 ($29 \pm 13$ $\mu$m/h) and primary fibroblast ($26 \pm 10$ $\mu$m/h) cells.
To extract the typical velocity of active nuclear movement, in each oscillatory 
period we determined its maximal speed. The distribution of these maximal velocities
(Fig.~6) shows a peak around $50 \mu$m/h. However, several nuclei were also
observed moving faster than 100 $\mu$m/h.

\paragraph{Microtubules are essential for nuclear migration.}

To determine the underlying mechanism driving the adhesion
constraint-induced nuclear motion, further experiments were carried out with
C6 cells in our micropatterned assay. Population-averaged nuclear speeds were
calculated as a quantitative measure of nuclear motion activity. The major
force generating cytoskeletal elements were targeted with specific drugs:
cytochalasin D (actin polymerization inhibitor), vinblastine (MT destabilizing
agent), taxol (MT stabilizator), ML-7 (myosin-II MLCK inhibitor), vanadate
(dynein inhibitor) and AMP-PNP (as kinesin inhibitor with low specificity).

Nuclear motion was strongly attenuated by both drugs that disrupt MT
dynamics (Table 1, Fig. 7, Video 3 and Video 4).  Taxol and vinblastine effectively blocked the
motion of the nuclei above 10 nM and 20nM concentrations, respectively. These
two drugs, however, have an opposite effect on cell morphology: while
vinblastine reduces, taxol increases the elongation of the cell shape. After the administration of taxol, several cells could divide, which confirms the mildness of the intervention. Effect of vinblastine was more massive: cells stopped nuclear migration abruptly. Although cells lost their long processes, they remained bipolar, which suggests a moderate disassembly of microtubules.

Drugs affecting the actin cytoskeleton and the actin-myosin system altered
nuclear motion only when administered at very high (above specific)
concentrations. In particular, 500 nM cytochalasin D reduced nuclear motion without  changing the cell shape (Video 5), at lower concentrations it was ineffective.
Kinesin and cytoplasmic dynein inhibitors had no effect on nuclear motility.
Finally, the replacement of the fibronectin-fibrinogen substrate to
poly-l-lysine did not alter nuclear motion, either.

\paragraph{MTOC position is correlated with the direction of nuclear motion.}

At the end of time-lapse recordings the cultures were fixed for anti-$\alpha$
tubulin and anti-$\gamma$ tubulin immunolabeling.  C6 cells, in which the
nucleus was moving just prior the fixation, were selected from the time-lapse
recordings. Subsequently, microtubules and MT organizing centers (MTOC) were
visualized in these cells by epifluorescence and confocal microscopy.  The MTOC
was always located either behind or beside the nucleus, between the plasma and
nuclear membranes, i.e., MTOC positions segregated into two groups, one group at the back, another group at the mid-point of the nucleus (Fig. 8). Generally, MTOC was positioned at the back of the nucleus, when the cell was fixed soon after the nucleus reversed its direction, and in cells with a nucleus at the middle of the cell, MTOC was found at the mid-point of the nucleus. Which means that those nuclei with MTOC at the mid-point moved faster.  Nuclei were not moving in cells exhibiting two
opposite MTOCs, each facing a different cell
process.  The anti-$\alpha$ tubulin immunostaining revealed densely packed microtubules,
mainly parallel to the cell axis (image not shown).

\newpage
\section*{Discussion}

\paragraph{Nuclear motion is needed for proper development. }

Active repositioning of the nucleus is an integral part of various cell
biological processes including locomotion and division \citep{Bray00}.  Most of
the {\it in vivo} occuring cases of nuclear motility has important
developmental aspects including fertilization, zygote formation,
cellularization and spatially and temporally regulated cell production.  The
most relevant example of oscillatory nuclear movement, interkinetic nuclear
migration, seems to be essential in the proper timing and abundancy of neuron
formation: The normal course of interkinetic nuclear migration can be blocked
by a defective function of the human LIS1 gene. In such cases a severe disease,
lissencephaly, develops which results in mental retardation, epilepsy and
usually death at an early age \citep{DT95,Morris00}.

\paragraph{Nuclear migration assay. }

Capitalizing on the promising potential of micropatterned surfaces in
biomedical engineering \citep{ZYAL99,FT00}, we developed an assay where cell
shape and migratory activity can be systematically controlled
\citep{BCH03}.  In particular, the narrow stripes of
attachment-permitting surfaces effectively inhibited cellular locomotion,
presumably as a result of hindered lamellipod formation.  This suppression of
locomotion helped us to uncouple the motion of the nucleus from that of the
whole cell.  Nevertheless cellular polarization, which we determined on the
bases of MTOC position as well as membrane ruffling distribution along the cell
perimeter, remained present.  The resulting assay thus recapitulates most
features of nuclear migration (cell polarization, reversive nuclear movement),
while allows the systematic study of a large number of individual cells. Our
experiments yielded the first direct evidence of attachment constraint-induced
reversive nuclear motion in vertebrate cells.

\paragraph{Nuclear positioning is MT-dependent.}

Nuclear migration was blocked by low concentrations of vinblastine or taxol.
These drugs are tubulin-specific, either depolymerizing or stabilizing MTs and
thus inhibiting their normal dynamics. Therefore, a functioning MT system is
essential for nuclear migration in our essay.  This finding is in concert with
results of recent studies performed in a variety of experimental systems, which
clearly indicated the fundamental role of microtubules in nuclear positioning
\citep{Morris00,Morris03}.  The critical role of a perinuclear microtubule fork 
in nuclear translocation has been revealed in neurons \citep{serine732}. In vitro experiments with purified proteins
demonstrated that microtubule polymerization alone is capable of pushing
MTOC-like structures into the geometrical center of microfabricated chambers
\citep{MTchamber1, MTchamber2}.  The force generated by tubulin polymerization
can be in the order of a few pN -- similar to that of exerted by molecular
motors \citep{DY97}.  MT polymerization and physical pushing of the nucleus was
demonstrated to contribute to nuclear positioning in fission yeast \citep{TMDIC01}.

In animal cells, gene products that affect MT stability like Lis1
\citep{SER97} and CLASP \citep{AHDS01} were uniquely found on MT plus ends.
The role of Lis1 in nucleokinesis was directly demonstrated by altered
migratory behavior of cerebellar granule cells explanted from a Lis1 knockout
mouse \citep{HFGB98}.  A number of other proteins like dynactin and CLIP
\citep{VTFEV99}, which are known to be involved in nuclear positioning of
mammalian cells, also specifically localize on MT plus ends.  Nevertheless, the
functional role of these proteins, as well as the operational integration of
this multi-component process remains unclear.  Experiments using vanadate and
the ATP analogue AMP-PNP as inhibitory agents yielded negative results, which
suggests that within our assay cytoplasmic dyneins and kinesins do not have a
central role in nuclear migration.  F-actin disruption by cytochalasin D and
myosin-II inhibition by the MLCK inhibitor ML-7
shows that nuclear migration is not sensitively actin- or myosin-dependent,
either.

\paragraph{Cell polarization.}

In accord with other observations on the directionality of cell movements and  MTOC  positioning, we found that cell polarization strongly correlated with the direction of nuclear movement. The MTOC was always located either behind or beside the nucleus, which reinforces a more complex relationship between MTOC position and the direction of cell movement than the general assumption that MTOC leads the way \citep{MTOCpositionsubstratum, centrosomesincellmovement, centrosome_directionality, centrosomeposition}. Moreover, the change in nuclear movement direction involved an immediate destabilization and partial collapse of the former leading edge. Further experiments may reveal the molecular connection between the change in cell polarization and the gradually decreasing nuclear velocity at the reversing points.

\paragraph{Cell cycle.}

The most prominent connection between cell cycle and oscillatory nuclear
movement is exhibited in the interkinetic nuclear migration of 
avian and mammalian neural stem cells \citep{TTC96}.  In this case the
cell cycle and the nuclear oscillation is so strongly coupled, that typically
there is a single oscillation in each cell cycle.  In fission yeast,
oscillatory nuclear motion was also observed during the prophase of meiosis
\citep{yeastnucloscill}.  In that case the cell shape was also strongly
elongated, however, the period of the motion was much shorter (ca. 10 minutes)
than the duration of the cell cycle (and even that of the prophase), allowing
multiple nuclear oscillations per cell cycle.  In our assay the period of the
reversive nuclear motility was substantially longer, in the range of 2-5 h.
This period is still much shorter than the typical cell cycle length,
approximately a day. Moreover, nuclear oscillations could be observed
practically during the entire cell cycle, not just prior mitosis.  Therefore,
in our assay we have no evidence for a strong coupling between cell cycle
phases and nuclear positioning.

\paragraph{Forces involved in nuclear migration.}

Forces associated with moving the nucleus in a viscous medium can be estimated
from the following considerations \citep{MTchambertheory}.  Cell plasma is not
a homogeneous liquid  as fibrous elements of the cytoskeleton have a
significant impact on the drag sensed by the nucleus.  Nevertheless, based on
the Stokes formula, an effective viscosity value $\eta$ can be obtained from
simultaneous measurements of the force $F$ acting on an intracellular passive
test particle, its size $r$ and velocity $v$, as $F = 6\pi\eta rv$. Such
measurements resulted in $\eta \approx 100$ Pas
\citep{viscosity,diffusionincells}.  Applying this result to the nucleus
(r=5$\mu$m, $v\approx50\mu$m/h) yields  $F\approx 100 pN$ for the driving
force.  We regard this value as an upper estimate, as it is highly likely that
cytoskeletal components are disassembled in front of the moving nucleus, thus
the apparent viscosity is less than that in the case of an inert test object.
Although currently it is impossible to estimate the significance of
cytoskeletal disassembly, we expect the driving force to be $F\approx 10 pN$ .
As 1 pN is the typical yield of molecular force generating mechanisms
(molecular motors, tubulin polymerization, etc.), the nuclear motile force
could be resulted by already a handful of molecular complexes.

\paragraph{Model for nucleokinesis.}

Based on our experiments we suggest that the force, which drives nuclear
migration in elongated cells, is MT-dependent, and originates from MT
polymerization.  At the observed nuclear migration speeds, the force a single
polymerizing MT can exert is $\sim 1$pN \citep{DY97}, thus in the
proposed model a few dozen MTs are expected to push the nucleus.  For a
steady, unidirectional motion the underlying left-right symmetry of the system
has to be broken. Usually this is realized by molecular
concentration gradients, or as in case of molecular motors, by the molecular
polarity of the substrate.

In our case a plausible explanation for the origin of symmetry breaking is
provided by the enforced narrow shape of the cell. In this configuration 
the microtubules emanating from the MTOC can preferably grow only in one 
direction (Fig.~9). As a result, the reactive force of their polymerization
due to dynamic instability
propels the MTOC together with the nucleus. When the nucleus encounters
a larger resistance or becomes ``stuck'' at the ``dead end'' of the
elongated cell, the force continues to act upon the MTOC and makes its former
location behind the nucleus unstable. As the MTOC is pushed ``over'' the
nucleus, gradually the MTOC-attached MTs are destabilizing while new ones are
growing towards the closer cell edge, resulting in a new, backward-directed 
propulsive force.

The asymmetry needed for directional nuclear movement could also be generated
by different MT polymerization rates at the front and rear cell process.
The rate difference could be resulted by the presence of a molecular gradient, 
strongly coupled to cell polarization and motility.
Such a gradient is exhibited, e.g., by the rho family GTPases, which are
considered as constituents of a master controller of directional cell
motility\citep{WW01} and their downstream targets like PKC$\zeta$ \citep{EH03}.
In this framework, a feedback is predicted which makes the maintenance of the
gradient incompatible with certain nuclear positions.  The conjectured feedback 
may involve the interaction of nuclear membrane-localized proteins with cdc42 
or its downstream targets.

\newpage
\section*{Conclusions}

C6 glioma cells, where reversive nuclear motion was most prominently exhibited,
share many properties with the radial glial cells, which were found
to be neural stem cells \citep{FBSS98, Campbell}.  It has also
been shown recently that microtubules, but not microfilaments, are critical for
the polarized morphology of the C6 cells \citep{LBHG03}.  However, as reversive
nuclear motion was not restricted to glia-derived cells, we presume the
phenomenon is the result of a general cell biological mechanism, present in
various cell types.  Although the clarification of this point needs further
studies, we predict that the elongated cell shape and an asymmetrically
localized MTOC should lead to the directed motion of the attached nucleus in
most cell types. 
\newpage
\section*{Acknowledgements}
The authors are most grateful to J\'anos Kov\'acs and Bal\'azs Heged{\H u}s for
their help in implementing the micropatterned technology. We thank Zs\'ofia
Jur\'anyi, Gergely Nagy and Rita Vass for their valuable help in  evaluating
the time-lapse recordings.  The useful comments of profs. J\'anos Kov\'acs and Dennis Bray are greatly appreciated. The mouse primary
fibroblasts were kindly provided by K\'aroly Mark\'o. We thank Emilia 
Madar\'asz for providing the conditions for cell culturing and 
immunocytochemistry. This research was
supported by Hungarian Science Research Funds, OTKA T034995 and F038110, and NKFP 3A/0005/2002.
\newpage
\bibliography{cza,phd}
\newpage
\suppressfloats
\clearpage

\begin{figure}
\caption{
  C6 cells on micropatterned surfaces. A phase contrast image is superimposed
  upon a background showing epifluorescence from 20 $\mu$m wide
  FITC-poly-l-lysine stripes (green). Due to the constrained cell attachments,
  cells assume elongated, bipolar morphology. The positions of cell nuclei 
  (arrowheads) are clearly recognizable. (Corresponding supplemental 
  material: Video 1.)
}
\end{figure}

\begin{figure}
\caption{ 
  Attachment constraint-induced reversive nuclear motion.  A well-spread C6
  cell (left panel) shows no nuclear displacements. The same cell type on the
  same substrate display reversive nuclear movement, when forced into a
  polarized morphology by seeding onto a micropatterned substrate (middle
  panel). Similar behavior is seen in primary fibroblast cultures (right
  panel), where the period of oscillation is decreased compared to that of C6
  cells. Images were taken 1h and 10 min apart for the C6 and the fibroblast
  cell, respectively. Scale bars: 50 $\mu$m. (Supplement showing
  the reversive nuclear motion of a C6 cell: Video 2.)
}
\end{figure}

\begin{figure}
\caption{ 
  Nuclear motion after cell divison. The reversive nuclear motion starts
  immediatelly after the reattachment of daughter cells, when the two nuclei
  begin to move in opposite directions.  The semi-synchronous cells do not seem
  to keep a continuous physical contact: cell-cell connections may break when
  the nuclei are at distal positions.  Scale bar: 100 $\mu$m.
}
\end{figure}

\begin{figure}
\caption{ 
 Nuclear oscillation period lengths in C6 (a) and primary fibroblast
 (b) cells. 
}
\end{figure}

\begin{figure}
\caption{ 
 Dynamics of reversive nuclear motion. 
 For selected cells the nucleus and the ends of both cell process were 
 followed frame-by-frame. The resulting positions -- along the stripes --
 are plotted versus the time elapsed after seeding in panel (a). The
 calculated corresponding nuclear velocities are shown in panel (b). During the
 20h long time period, the traced nucleus completed 4 full oscillations. The
 gradually increasing cell length reflects cell reattachment after a division.
 After each direction reversal, the former leading edge partially detaches
 (arrowheads).  The direction reversal of the nucleus occurs within a process
 which changes the nuclear velocity in an almost linear fashion.
 The dotted lines mark the time moments, when the nucleus moves with a 
 maximal speed.
}
\end{figure}

\begin{figure}
\caption{ 
 The distribution of the
 maximal velocities within each period reveals that the typical 
 speed of active nuclear movement is around 50 $\mu$m/h.
}
\end{figure}

\begin{figure}
\caption{
  Cytoskeleton affecting drugs can inhibit nuclear migration in C6 cells.  The
  effects of drugs were characterized by the relative change in average nuclear
  speeds before and after the administration of the drug. Error bars represent
  standard deviation.  Both vinblastine and taxol had a robust inhibitory
  effect (p$<10^{-4}$), while the rest of the drugs did not result in a significant alteration
  of nuclear motion (p$>0.1$).
}
\end{figure}

\begin{figure}
\caption{ 
 Correlation between MTOC position and nuclear migration direction. MTOC 
 position was determined by $\gamma$-tubulin immunostaining of the specimen,
 fixed immediatelly after the recordings (panel a). The obtained MTOC positions
 are summarized in a shematic diagram (panel b). The arrow indicates the
 direction of motion. Cells had either one (full circles) or two (crosses)
 MTOCs. When the investigated nucleus was moving, the attached MTOC was {\em
 never} found in front of the nucleus.
}
\end{figure}

\begin{figure}
\caption{ 
 Schematic model of micotubule polimerization-driven nuclear migration.
 (a) The close association of the nucleus and microtubule organizing center 
 results in an asymmetric microtubule distribution. Forces arising at the
 growing microtubule tips are transmitted to the nucleus through the MTOC.  (b-d)
 When the motion of the nucleus is hindered, the force continues to act upon
 the MTOC and makes its former location behind the nucleus unstable. Thus, the
 MTOC is pushed in front of the nucleus. (e) As a result of dynamic instability,
 eventually new microtubules originate from the MTOC, yielding a 
 backward-directed propulsive force.
}
\end{figure}

\clearpage
\paragraph{Legend for Video 1.} 

Motility of C6 cells on the surface of 20 $\mu$m wide 
Alexa488-labeled cell adhesive fibronectin-fibrinogen stripes (green). Frames 
were collected in every 5 minutes. Display rate: 10 frames/second. Video 1 is
related to Fig. 1.

\paragraph{Legend for Video 2.}

Reversive nuclear motion in a single C6 cell on a 20 $\mu$m wide 
Alexa488-labeled cell adhesive fibronectin-fibrinogen stripe. Frame size: 270x60
$\mu$m$^2$. Frames were collected in every 5 minutes. Display rate: 8 
frames/second. Video 2 is related to Fig. 2.

\paragraph{Legend for Video 3.}

The effect of 30 nM taxol administered 16 h after the beginning of the movie on auto-reverse nuclear migration in C6 cells cultured on 20 $\mu$m wide stripes. Frames  were collected in every 5 minutes. Display rate: 8 frames/second. 

\paragraph{Legend for Video 4.}

The effect of 20 nM vinblastine administered 9 h after the beginning of the movie on auto-reverse nuclear migration in C6 cells cultured on 20 $\mu$m wide stripes. Frames  were collected in every 5 minutes. Display rate: 8 frames/second. 

\paragraph{Legend for Video 5.}

The effect of 500 nM cytochalasin D administered 12 h after the beginning of 
the movie on auto-reverse nuclear migration in C6 cells cultured on 20 $\mu$m 
wide stripes. Frames  were collected in every 5 minutes. Display rate: 10 
frames/second.

\clearpage
\newcommand{\rb}[1]{\raisebox{1.5ex}[0pt]{#1}}
%\begin{document}

\begin{table}[pt]
\centering
\begin{tabular}{|r|c|c|c|c|r|c|r|}
\hline
        &       &    &   & \multicolumn{2}{c|}{day 1}& \multicolumn{2}{c|}{day 2}\\
\rb{sample}& \rb{cell type} & \rb{substrate} & \rb{drug} &
	\multicolumn{1}{c}{velocity [$\mu$m/h]}&
	\multicolumn{1}{c|}{n}&
	\multicolumn{1}{c}{velocity [$\mu$m/h]}&
	\multicolumn{1}{c|}{n}\\ 
\hline\hline
1 	&  C6	& FN  & --		&$ 27\pm22	$&$ 871		$&$ 27\pm14	$&$ 757 	$\\ \hline
2 	&  C6	& FN  & --		&$ 28\pm13	$&$ 728		$&$ n.a. 	$&$ n.a.	$\\ \hline
3 	&  C6	& PLL & --		&$ 23\pm13	$&$ 841		$&$ 28\pm14	$&$ 347		$\\ \hline
4 	&  C6	& PLL & --		&$ n.a.		$&$ n.a.	$&$ 33\pm8	$&$ 965		$\\ \hline
5 	&  C6	& PLL & --		&$ 25\pm5	$&$ 1065	$&$ n.a.	$&$ n.a.	$\\ \hline
6 	&  C6	& PLL & --		&$ 21\pm11	$&$ 1123	$&$ 32\pm13	$&$ 431		$\\ \hline
7 	&  PMF  & PLL & --		&$ 26\pm9	$&$ 488		$&$ 24\pm6	$&$ 1314	$\\ \hline
8 	&  PMF  & PLL & --		&$ 15\pm3	$&$ 256		$&$ 24\pm16	$&$ 1051	$\\ \hline
9 	&  C6	& PLL & 100$\mu$M AMP-PNP&$ 31\pm10	$&$ 844		$&$ 34\pm8	$&$ 463		$\\ \hline
10 	&  C6	& PLL & 100$\mu$M AMP-PNP&$ 37\pm4	$&$ 476		$&$ 35\pm11	$&$ 374		$\\ \hline
11 	&  C6	& PLL & 200nM citochalasin D &$ 10\pm5	$&$ 407		$&$ 12\pm5	$&$ 114		$\\ \hline
12	&  C6	& PLL & 500nM citochalasin D &$ 17\pm6	$&$ 666		$&$ 10\pm3	$&$ 956		$\\ \hline
13	&  C6	& FN  & 1$\mu$M ML-7 	&$ 17\pm7	$&$ 879		$&$ 19\pm5	$&$ 674		$\\ \hline
14	&  C6	& FN  & 30 nM taxol 	&$ 25\pm11	$&$ 372		$&$ 8\pm2	$&$ 390		$\\ \hline
15	&  C6	& PLL & 5$\mu$M vanadate &$ 30\pm9	$&$ 357		$&$ 31\pm11	$&$ 378		$\\ \hline
16	&  C6	& PLL & 20nM vinblastine &$ 34\pm0	$&$ 306		$&$ 3\pm0	$&$ 262		$\\ \hline
\end{tabular}
\caption{
  Summary of experimental data.  Population-averaged nuclear velocities were
  obtained from cells engaing in reversive nuclear motion. $n$ denotes the
  number of positions used for calculationg the average velocities. PMF:
  primary mouse fibroblast; FN: Fibronectin-Fibrinogen; PLL: poly-l-lysine.
}
\end{table}
\vfill
%\end{document}

\end{document}